\documentclass[aps, prb, twocolumn, amssymb, amsmath, superscriptaddress]{revtex4-1}

\usepackage{bm}
\usepackage{times}
\usepackage{graphicx}
\usepackage[colorlinks,citecolor=blue,linkcolor=blue]{hyperref}
\usepackage{array}
\usepackage{float}
\usepackage{colortbl}
\usepackage{multirow}
\usepackage{tabularx}

\begin{document}

\title{Possible Superconductivity Approaching Ice Point}
\author{Yanfeng Ge}
\affiliation{School of Physics, Beijing Institute of Technology, Beijing 100081, China}
\author{Fan Zhang}\email{zhang@utdallas.edu}
\affiliation{Department of Physics, University of Texas at Dallas, Richardson, Texas 75080, USA}
\author{Yugui Yao}\email{ygyao@bit.edu.cn}
\affiliation{School of Physics, Beijing Institute of Technology, Beijing 100081, China}

\date{\today}

\begin{abstract}
Recently BCS superconductivity at 203~K has been discovery in a highly compressed hydrogen sulfide.
We use first-principles calculations to systematically examine the effects of partially substituting the chalcogenide atoms on the superconductivity of hydrogen chalcogenides under high pressures.
We find detailed trends of how the critical temperature changes with increasing the V-, VI- or VII-substitution rate,
which highlight the key roles played by low atomic mass and by strong covalent metallicity.
In particular, a possible record high critical temperature of 280~K is predicted in a stable H$_{\bm 3}$S$_{\bm{0.925}}$P$_{\bm{0.075}}$ with the $\bm{Im\bar{3}m}$ structure under 250~GPa.
\end{abstract}

\maketitle

\section{Introduction}

Since the discovery of superconductivity in mercury at $4$~K in 1911~\cite{Delft2010}, scientists have been ardently pursuing
new and better superconductors at higher temperatures.
In a conventional superconductor, the vibrations of crystal lattice provide an attractive force that binds an electron with its time-reversal partner into a Cooper pair~\cite{Bardeen1957}.
The Cooper pairs can Bose condense below a critical temperature ($T_c$), which had been believed to be no larger than $25$~K prior to the discovery of $\rm MgB_2$~\cite{Nagamatsu2001,Budko2001}.
In this case, the Debye temperature could be as large as $\sim 10^3$~K, and low $T_c$ is primarily related to small electron-phonon couplings~\cite{Ginzburg}.
Since 1986 and 2008, respectively, the discoveries of copper- and iron-based superconductors have provided two new avenues for making high-$T_c$ superconductors~\cite{Bednorz1986,Chu,Schilling1993,Gao1994,Kamihara2008,Ren2008,Wu2009},
while generating new excitements in fundamental physics.
Although their unconventional mechanisms are still under hot debate, to date a record $T_c$ of $164$~K
has been experimentally realized in the cuprate family~\cite{Gao1994} and $56$~K among the iron-pnictide compounds~\cite{Wu2009}.

One may thus wonder whether a $T_c$ higher than $164$~K could be achieved and whether the conventional phonon-mediated mechanism could play a significant role in such a race.
Both answers have become extremely positive and fascinating, given the recent theoretical prediction~\cite{Li2014,Duan2014}
and particularly the most recent experimental observation~\cite{Drozdov2014,Drozdov2015} of superconductivity at $203$~K in a highly compressed hydrogen sulfide~\cite{Li2014,Duan2014,Drozdov2014,Drozdov2015,Bernstein2015,Duan2015,Errea2015,Nicol2015,Papaconstantopoulos,Akashi,Heil}.
As pointed out by Bernstein {\it et al.} in a microscopic theory of $\rm H_3S$~\cite{Bernstein2015},
the substantial covalent metallicity leads to a large electron-phonon coupling,
and the low atomic mass leads to high-frequency phonon modes.
Both features play essential roles in increasing the $T_c$ of $\rm H_3S$.
Notably, the former feature is analogous to that of $\rm MgB_2$~\cite{An2001,Mazin2003} whereas the latter effect is similar to those in the H-rich materials \cite{Ashcroft1968,Ashcroft2004,Feng2006,Tse2007,Cudazzo2008,Gao2008,Eremets2008,Zureka2009,Kim2010,Kim2011,Flores-Livas2012,Wang2012}.

As a powerful method for the optimization of $T_c$, the atomic substitution has been routine in superconductivity experiments~\cite{Cava1990,Ekimov2004,Lv2008,Kimber2009}.
However, there is yet any study on how the atomic substitution influences the $T_c$ of $\rm H_3S$ and of the more general hydrogen chalcogenides.
Here, we systematically examine the effects of (partially) substituting the chalcogenide atoms on the superconductivity
and particularly the $T_c$'s of hydrogen chalcogenides in the $Im\bar{3}m$ phase under high pressures, based on the first-principles calculations with virtual crystal approximation (VCA)~\cite{Nordheim1931,Rosner2002,Boeri2004,Lee2004,Blackburn2011}.
In the V- and VII-substitution cases, we find that the significant changes of the DOS at Fermi surface and of the phonon linewidths,
coming from the different number of valance electrons, are the principal factors affecting the electron-phonon coupling and $T_c$.
In the particular case of $\rm H_3S_{1-x}P_{x}$ at $200$~GPa,
we find that the DOS, the electron-phonon coupling constant, and the $T_c$ all first increase and then decrease as the P-substitution rate increases from zero to $0.2$.
In the optimized condition in which $x=0.075$ and the pressure increases to $250$~GPa, a possible record high $T_c$ of $280$~K is predicted.
In contrast, the $T_c$ decreases monotonously with increasing the VII-substitution degree.
In the VI-substitution cases, the $T_c$ does not appear to increase substantially, because of the reduction of DOS with O- or Se-substitution, or due to the softening of logarithmically averaged phonon frequency with Te-substitution.
These findings emphasize the importance of low atomic mass and strong covalent metallicity in conventional high-$T_c$ superconductors,
pave the way for substantially enhancing $T_c$ by combining application of a high pressure and properly designed chemical substitution,
and suggest that in principle it is not impossible to boost $T_c$ to ice point in optimized conditions.

\section{Results}

\begin{figure*}[ht!]
\centerline{\includegraphics[width=\textwidth]{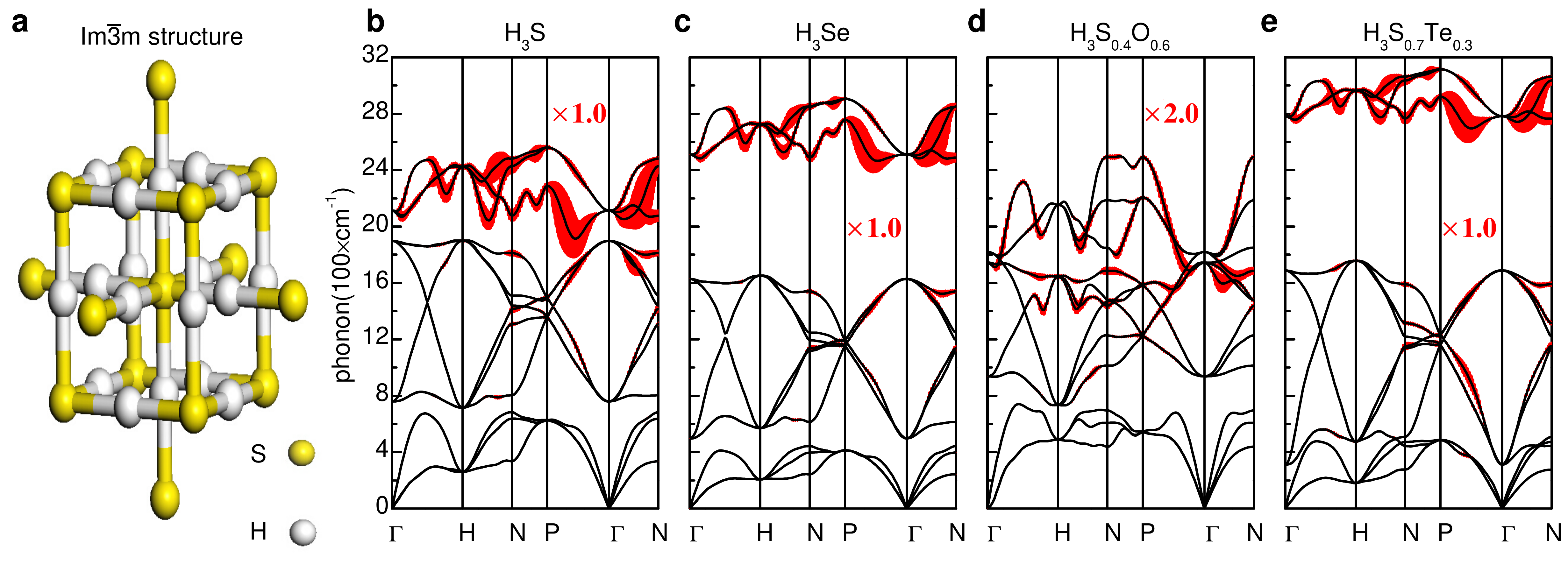}}
\caption{{\bf $\bm{Im\bar{3}m}$ structure of H$_{\bm 3}$S and phonon spectra with phonon linewidth of VI-substitution systems.}
(a) The $Im\bar{3}m$ structure of hydrogen chalcogenides.
(b)-(e) Phonon spectra and phonon linewidths of four representative VI-substitution systems:
(b) $\rm H_3S$, (c) $\rm H_3Se$, (d) $\rm H_3S_{0.4}O_{0.6}$, and (e) $\rm H_3S_{0.7}Te_{0.3}$.
The magnitude of the phonon linewidth is indicated by the size of red error bar, and the magnitude for $\rm H_3S_{0.4}O_{0.6}$ is plotted with twice of the real values.
\label{fig:6ph}}
\end{figure*}

Soon after the experimental report~\cite{Drozdov2014,Drozdov2015} on the record $T_c$ of a highly-compressed hydrogen sulfide~\cite{Li2014,Duan2014},
further theoretical studies~\cite{Bernstein2015,Duan2015,Errea2015,Nicol2015,Papaconstantopoulos,Akashi,Heil} have reached two consensuses:
(i) the superconducting matter is most likely $\rm H_3S$ with a strong covalent character in the $Im\bar{3}m$ phase, as sketched in Fig.~\ref{fig:6ph}a,
and (ii) the superconductivity of $\rm H_3S$ is phonon-mediated. Thus, the superconductivity of highly-compressed hydrogen sulfide can be accurately described by the celebrated Eliashberg theory~\cite{Allen1983},
which takes into account the renormalization of electron-electron repulsion by electron-phonon interactions.
In this theory, the Allen-Dynes-modified McMillan formula~\cite{Allen1975,Durajski} relates $T_c$ to the logarithmically averaged phonon frequency $\omega_{\ln}$, the effective Coulomb repulsion $\mu^\ast$,
and the electron-phonon coupling constant $\lambda$:
\begin{eqnarray}
{{T_c}}&=f_1f_2&\frac{{\omega}_{\ln}}{1.20}
\exp\left[-\frac
{1.04(1+\lambda)}
{\lambda-\mu^\ast(1+0.62\lambda)}\right]\,,
\label{eq:tc}
\end{eqnarray}
where $f_1$ and $f_2$ are the strong coupling and the shape correction factors~\cite{Allen1975}, respectively.
Clearly, there are two ways to enhance $T_c$, i.e., to search for materials with high-frequency phonon modes and to increase electron-phonon couplings.
As mentioned in the introduction, the $\rm H_3S$ happens to constitute both advantages and hence has a record high $T_c$.
Indeed, within this framework recent theoretical calculations have well explain the high $T_c$ in the highly-compressed hydrogen sulfide~\cite{Li2014,Duan2014,Drozdov2014,Drozdov2015,Bernstein2015,Duan2015,Errea2015,Nicol2015,Papaconstantopoulos,Akashi,Heil,Quan}.
Therefore, we will first reproduce the $T_c$ of $\rm H_3S$ in the $Im\bar{3}m$ phase at $200$~GPa and study the substitution of $\rm S$ atoms by other VI atoms.
Comparing our results with previous ones will show the validity of our first-principles calculations with VCA.

In the VI-substitution systems at $200$~GPa, we focus on studying $\rm H_3S_{1-x}Se_{x}$ in the full range of $x$,
$\rm H_3S_{1-x}O_{x}$ in the range of $x=0.0 \sim 0.6$, and $\rm H_3S_{1-x}Te_{x}$ in the range of $x=0.0 \sim 0.3$.
Because when $x$ takes a greater value in the case of O- or Te-substitution,
we can find imaginary phonon modes at $\Gamma$ point, indicating a lattice structure instability.
The electronic band structures of these systems are shown in the supplementary materials,
and our results for $\rm H_3S$ and $\rm H_3Se$ are in great agreement with previous reports based on similar first-principles calculations~\cite{Duan2014,FloresLivas,Zhang}.
We find that the VI-substitutions have little influence on the electronic band structure around the Fermi energy, except unimportant changes near $\Gamma$ point (see supplementary materials).
We further find that including spin-orbit couplings for the Te-substitution case does not introduce any significant correction, either.
Notably, the main electronic effect is the reduction of DOS at Fermi surface in all of these VI-substitution cases, as summarized in Table~\ref{tab:dos}.
The DOS decreases monotonically with increasing the VI-substitution rate. Moreover, the lighter the substitution element, the stronger the reduction of DOS.

\begin{table}[b!]
\caption{\bf Fermi-surface DOS (in units of Hartree$^{\bf -1}$/spin) of hydrogen chalcogenides in various V-, VI-, and VII-substitution systems at $\bf 200$ GPa.}
\begin{tabular*}{8.5cm}{@{\extracolsep{\fill}}cccccccc}
\hline\hline  & $x$ &  H$_3$S$_{1-x}$O$_{x}$  &  H$_3$S$_{1-x}$Se$_{x}$ &  H$_3$S$_{1-x}$Te$_{x}$ \\
\hline &  0.0& 7.91  & 7.91 & 7.91 \\
       &  0.1& 7.70  & 7.78 & 7.90 \\
       &  0.3& 6.84  & 7.61 & 7.53 \\
       &  0.6& 4.53  & 7.51 &      \\
       &  1.0&       & 7.31 &      \\
\hline\hline  & $x$ &  H$_3$S$_{1-x}$P$_{x}$  &  H$_3$Se$_{1-x}$As$_{x}$  & H$_3$S$_{1-x}$Cl$_{x}$  & H$_3$Se$_{1-x}$Br$_{x}$  \\
\hline &  0.000& 7.91  & 7.31 & 7.91 & 7.31 \\
       &  0.025& 8.57  & 7.80 & 7.18 & 6.50 \\
       &  0.050& 9.11  & 7.79 & 6.58 & 5.85 \\
       &  0.075& 9.92  & 7.90 & 5.63 & 5.23 \\
       &  0.100&10.55  & 7.96 & 5.02 & 4.72 \\
       &  0.125& 9.21  & 8.27 & 4.84 & 4.55 \\
       &  0.150& 8.02  & 8.38 & 4.71 & 4.39 \\
       &  0.175& 7.22  & 8.12 & 4.59 & 4.25 \\
       &  0.200& 6.69  & 6.92 & 4.47 & 4.14 \\
\hline \hline
\end{tabular*}
\label{tab:dos}
\end{table}

\begin{figure*}[t!]
\centerline{\includegraphics[width=\textwidth]{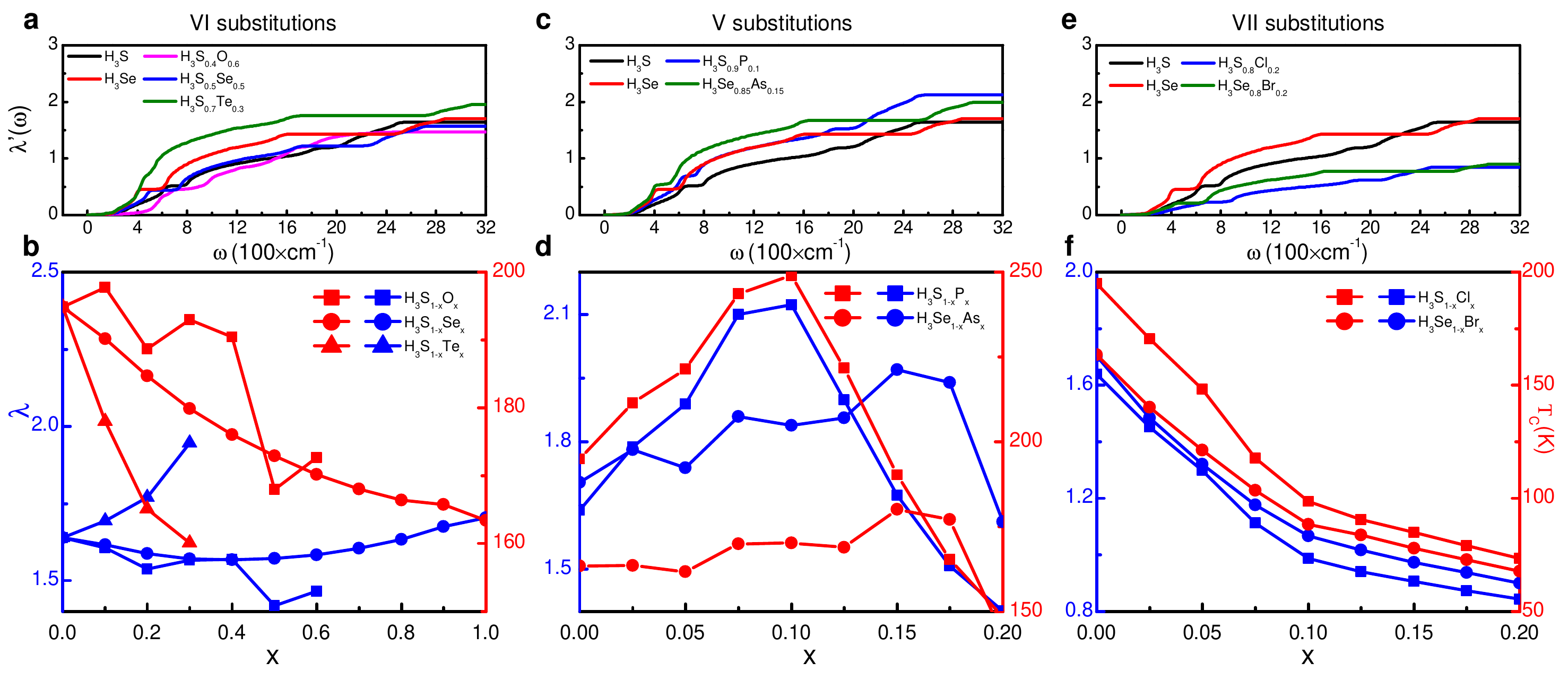}}
\caption{{\bf Electron-phonon coupling and critical temperature at 200~GPa of hydrogen chalcogenides with chemical substitutions.}
Upper panels: the integral function $\lambda'(\omega)$.
Lower panels: the electron-phonon coupling constant $\lambda$ (solid lines) and the critical temperature $T_c$ (dashed lines) versus the substitution concentration $x$.
The left, middle, and right panels respectively show the results for the case of VI-, V-, and VII-substitutions.
\label{fig:tc}}
\end{figure*}

Figures~\ref{fig:6ph}b-\ref{fig:6ph}e display the phonon spectra of four representative VI-substitution systems,
and the magnitude of the phonon linewidth is indicated by the size of the red error bar.
Our results for $\rm H_3S$ and $\rm H_3Se$ are again consistent with the earlier reports~\cite{Errea2015,Zhang}.
Overall there is a clear separation between H modes at high energy and S modes at low energy~\cite{Errea2015}.
Evidently, with increasing the degree of heavy-element (Se or Te) substitution, all the three acoustic phonon modes decrease in frequency.
In addition, the H-VI bond-bending modes (displacement of an H atom perpendicular to a H-VI bond) in the intermediate frequency region are softened,
whereas the H-VI bond-stretching modes (displacement of an H atom parallel to a H-VI bond) in the high frequency region are hardened.
Not surprisingly, the opposite trends occur for the case of O-substitution.
The above changes at low and intermediate frequency are expected by considering the relative atomic mass of the substitution elements,
whereas the changes at high frequency might be explained by the dependence of the chemical precompression effect~\cite{Ashcroft2004,Zhang,Zhong} on the atomic radius of VI elements.

Further analyzing phonon linewidths reveals that they are larger for the H vibrational modes, particularly for the bond-stretching modes at high frequency,
similar to the results by Errea {\it et al.}~\cite{Errea2015}
We note that the small magnitude of phonon linewidth in $\rm H_3S_{0.4}O_{0.6}$ (Fig.~\ref{fig:6ph}d) is ascribed to the reduction of DOS.
In order to expose more clearly the relative contributions of different phonon modes to the electron-phonon coupling constant $\lambda$,
we define an integral function $\lambda'(\omega) = 2\int_0^\omega\omega'^{-1}\alpha^2F(\omega')\rm{d}\omega'$,
where $\lambda'(\infty)$ is $\lambda$ and $\alpha^2F(\omega)$ is the Eliashberg function~\cite{McMillan1968,Allen1972,Allen1974}.
As shown in Fig.~\ref{fig:tc}a, the contribution from the intermediate frequency region ($300\sim 1700$~cm$^{-1}$) has an increase in $\rm H_3S_{0.7}Te_{0.3}$,
which results in a corresponding increase in the electron-phonon coupling constant.
Such a variation is a consequence of the increase of phonon linewidths in the intermediate frequency region,
which is present in Fig.~\ref{fig:6ph}e but absent in the O- and Se-substitution systems (Figs.~\ref{fig:6ph}c-\ref{fig:6ph}d).

With the above results, we use equation~(\ref{eq:tc}) to estimate the $T_c$ for the investigated VI-substitution systems and the results are plotted in Fig.~\ref{fig:tc}b.
The effective Coulomb repulsion has been chosen to be $\mu^\ast=0.12$, in the reasonable range of $0.1\sim 0.15$~\cite{Grimvall},
such that the estimated $T_c$ for $\rm H_3S$ and $\rm H_3Se$, $194$~K and $160$~K, are most close to the reported values~\cite{Errea2015,FloresLivas}.
In all of the VI-substitution cases, there is no substantial enhancement of $T_c$ compared with the $194$~K in $\rm H_3S$,
except a slightly higher $T_c$ of $198$~K in $\rm H_3S_{0.9}O_{0.1}$.
Notably, despite a higher $\lambda$ in $\rm H_3S_{0.7}Te_{0.3}$, the $T_c$ is in fact reduced to $161$~K.
Because of the heavy mass of Te, the logarithmically averaged phonon frequency $\omega_{\ln}$ is only $90.8$~meV in $\rm H_3S_{0.7}Te_{0.3}$, much lower than the $125.2$~meV in $\rm H_3S$.
In fact, the effect of VI-substitution at a fixed physical pressure can be viewed as a chemical pressure effect.
Thus, our results that the $T_c$ would decrease in the various VI-substitution cases at $200$~GPa are in agreement with
the experimental observation~\cite{Drozdov2015} and an earlier theoretical study~\cite{Akashi}, which have confirmed that the $T_c$ of $\rm H_3S$ is peaked near $200$~GPa.

Besides the substitution of $\rm S$ by elements in the same main group, we also further consider substitutions by the adjacent group elements, i.e., phosphorus and halogen groups.
We mainly focus on the cases in which a small percentage ($x=0.0\sim0.2$) of S is replaced by  P (Cl), and likewise Se replaced by As (Br).
In sharp contrast to the VI-substitution cases, the averaged phonon frequency is not expected to have a significant drop after substitutions,
because the atomic mass of after chemical substitutions are close to that of the prototypes.
Thus, as the calculations will show below, the changes of $T_c$ follow closely the changes of $\lambda$ in the V- and VII-substitution cases.
Note that the absence of imaginary frequency in the phonon spectra, as exhibited in Fig.~\ref{fig:57ph} (also see supplementary materials),
ensures the lattice dynamic stabilities for all the cases of interest.

\begin{figure*}[t]
\centerline{\includegraphics[width=0.8\textwidth]{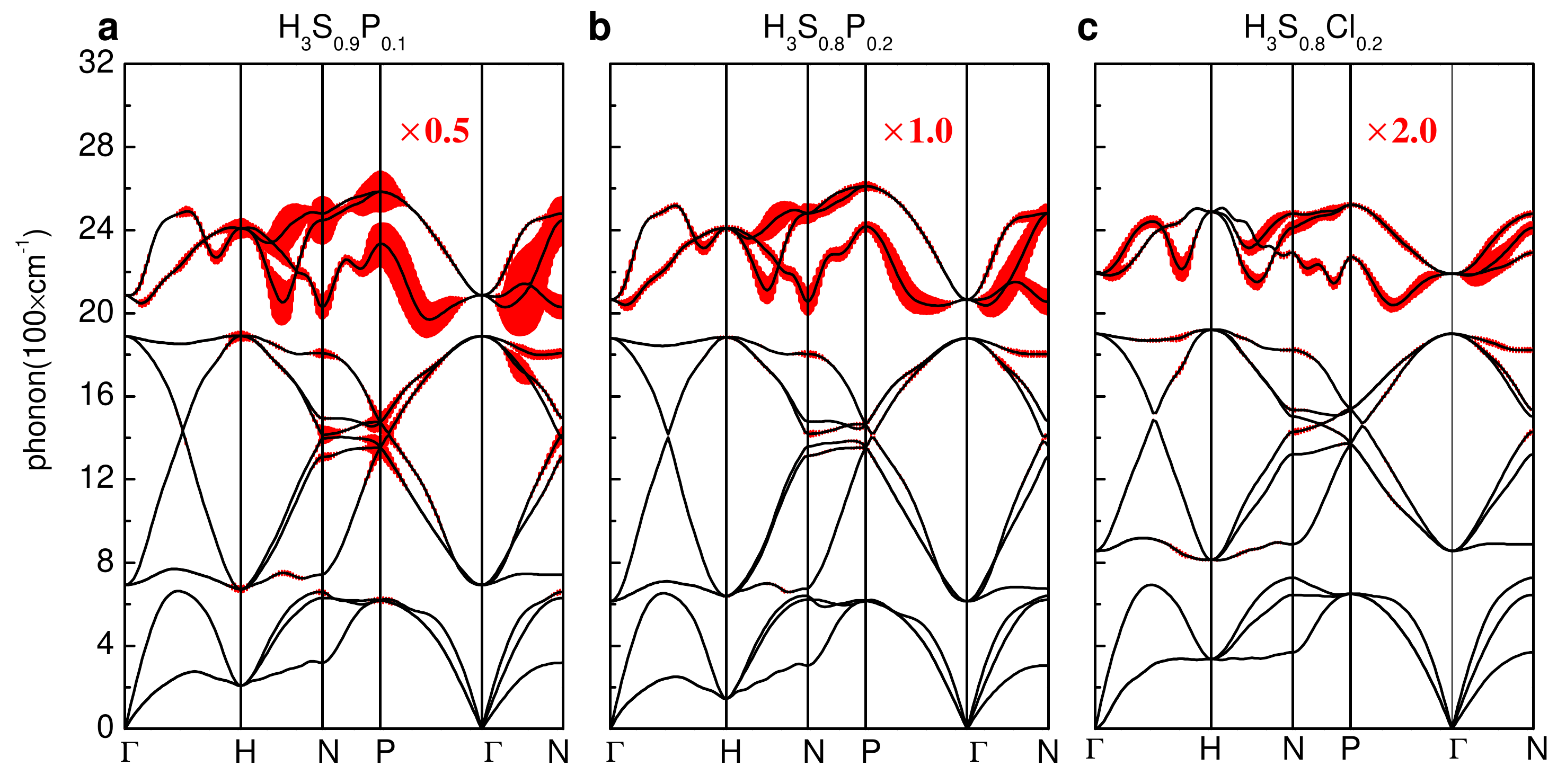}}
\caption{{\bf Phonon spectra with phonon linewidth of V- and VII-substitution systems.}
(a) $\rm H_3S_{0.9}P_{0.1}$, (b) $\rm H_3S_{0.8}P_{0.2}$, and (c) $\rm H_3S_{0.8}Cl_{0.2}$.
The magnitude of phonon linewidths in $\rm H_3S_{0.9}P_{0.1}$ ($\rm H_3S_{0.8}Cl_{0.2}$ ) is too large (small) and thus plotted with half (twice) of the real values.
\label{fig:57ph}}
\end{figure*}

Due to the decrease of valence electrons after V-substitutions, the electronic energy bands shift up a little bit with respect to the Fermi energy (see supplementary materials).
Such changes can also be reflected effectively by the changes in DOS at the Fermi surface, as summarized in Table~\ref{tab:dos}.
It turns out that the DOS increase first and then decrease with increasing the V-substitution rates.
Notably, the DOS can reach very large values, e.g., $10.55$ in $\rm H_3S_{0.9}P_{0.1}$ and $8.38$ in $\rm H_3Se_{0.85}As_{0.15}$, in units of Hartree$^{-1}$ per spin.
Although the phonon spectra hardly change after V-substitutions, the phonon linewidths follow a close trend of the changes in DOS.
The magnitudes of phonon linewidths in the intermediate and high frequency regions rise sharply as the P-substitution rate reaches $x=0.1$
and then fall steeply as the rate further increase to $x=0.2$, as shown in Figs.~\ref{fig:57ph}a-\ref{fig:57ph}b.
Thus, the electron-phonon coupling constant $\lambda$ follows the same trend of the changes in DOS and in phonon linewidths,
as shown in Fig.~\ref{fig:tc}d.
Analysis of $\lambda'(\omega)$ plotted in Fig.~\ref{fig:tc}c further confirms the enhancement of electron-phonon couplings
in the intermediate and high frequency regions in $\rm H_3S_{0.9}P_{0.1}$.
Very similar behaviors can be found as the As-substitution rate increases from zero to $x=0.2$ in $\rm H_3Se_{1-x}As_{x}$,
as seen in Figs.~\ref{fig:tc}c-\ref{fig:tc}d (also see supplementary materials).
We find that the maximal coupling constant values are $2.12$ in $\rm H_3S_{0.9}P_{0.1}$ and $1.97$ in $\rm H_3Se_{0.85}As_{0.15}$.

Based on the above results, we use equation~(\ref{eq:tc}) with $\mu^\ast=0.12$ to estimate $T_c$ for the two V-substitution systems and plot them in Fig.~\ref{fig:tc}d.
The $T_c$ are found to be as high as $250$~K in $\rm H_3S_{0.9}P_{0.1}$ and $185$~K in $\rm H_3Se_{0.85}As_{0.15}$,
which are greatly enhanced from the $194$~K ($160$~K) in the prototypical $\rm H_3S$ ($\rm H_3Se$).
Given the possible variation of the effective Coulomb repulsion $\mu^\ast$ in the range of $0.1\sim0.15$,
the $T_c$ in $\rm H_3S_{0.9}P_{0.1}$ may also vary in the range of $227\sim265$~K.

\begin{figure*}[t]
\centerline{\includegraphics[width=0.7\textwidth]{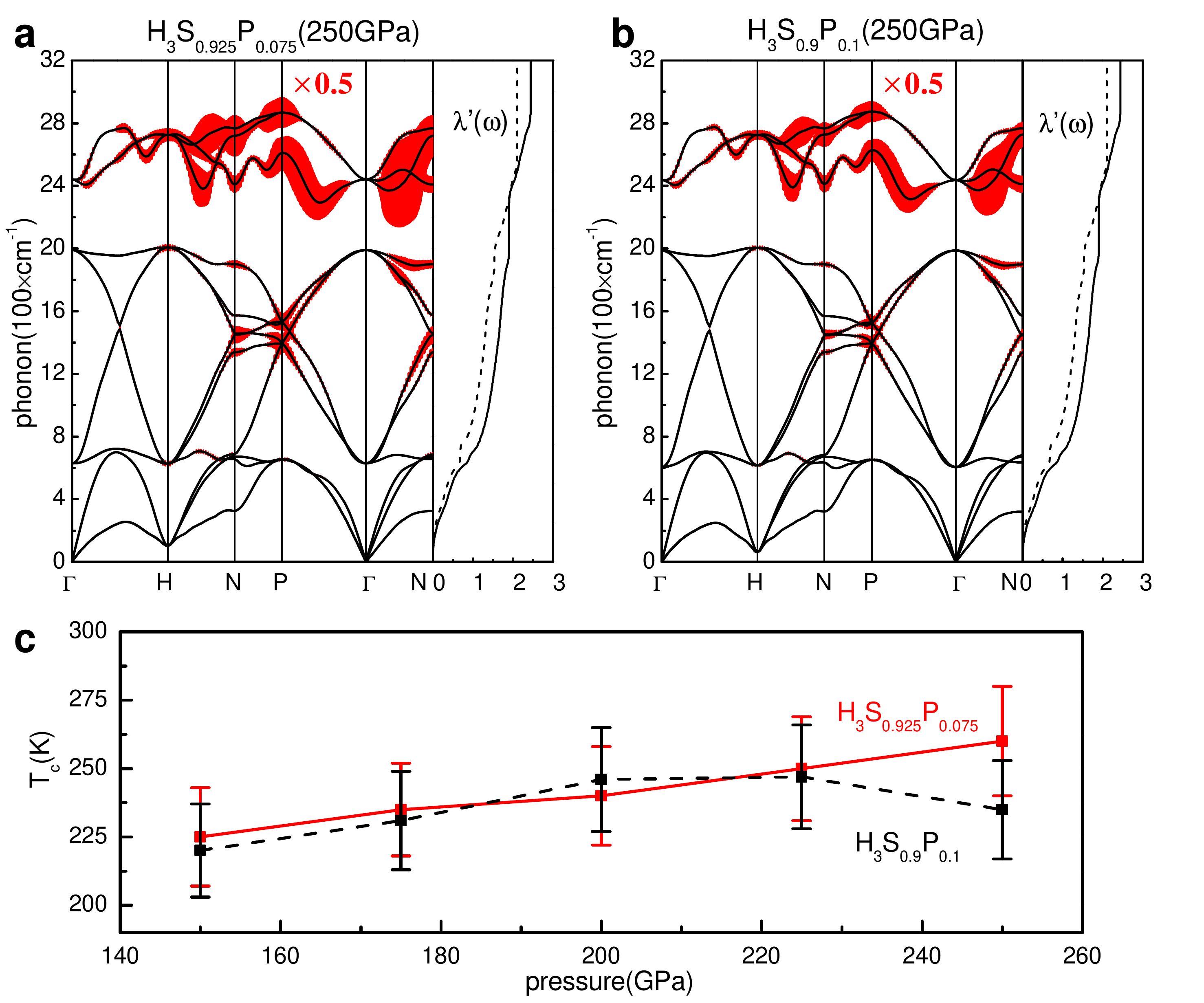}}
\caption{{\bf The effect of pressure in $\rm \bf H_3S_{1-x}P_{x}$.}
(a)-(b) The phonon spectra and $\lambda'(\omega)$ of $\rm H_3S_{0.925}P_{0.075}$ and $\rm H_3S_{0.9}P_{0.1}$ at $250$~GPa.
The magnitude of phonon linewidth is plotted with half of the real values.
The solid (dash) lines plot $\lambda'(\omega)$ at $250$ ($200$)~GPa.
(c) $T_c$ of $\rm H_3S_{0.925}P_{0.075}$ (red solid line) and $\rm H_3S_{0.9}P_{0.1}$ (black dash line) versus pressure.
The error bars indicate the value range of $T_c$ with $\mu^\ast=0.1\sim0.15$.
\label{fig:press}}
\end{figure*}

We now take into account the influence of varying the high pressure on the superconductivity~\cite{Drozdov2014,Drozdov2015} in the case of $\rm H_3S_{1-x}P_{x}$.
Since we are particularly interested in optimizing the $T_c$, we focus on two cases suggested by Fig.~\ref{fig:tc}d: $\rm H_3S_{0.9}P_{0.1}$ and $\rm H_3S_{0.925}P_{0.075}$.
The main results are summarized in Table~\ref{tab:pressdos} and Fig.~\ref{fig:press}.
In the case of $\rm H_3S_{0.9}P_{0.1}$, the $T_c$ reaches the highest value at $200$~GPa and remains the same value up to $225$~GPa,
whereas in the case of $\rm H_3S_{0.925}P_{0.075}$, the $T_c$ increases monotonously as the pressure rises from $150$~GPa to $250$~GPa, beyond which a structure instability is found.
Results in the latter case follow the fact that the DOS at the Fermi energy, the phonon linewidths, and the electron-phonon coupling constants all increase gradually with increasing the pressure.
For $\mu^\ast=0.1$, the $T_c$ of $\rm H_3S_{0.925}P_{0.075}$ at $250$~GPa can reach $280$~K, which is higher than the ice point.
We note that the pressure of $250$~GPa has already been feasible in the $\rm H_3S$ experiment by Drozdov {\it et al}~\cite{Drozdov2015}.
Given the range of $\mu^\ast=0.1\sim0.15$, the $T_c$ in such an optimized case is at least as high as $241$~K, as indicated by the error bar in Fig.~\ref{fig:press}c.
(For $\mu^\ast=0.12$ that yields the $T_c$ of $194$~K in $\rm H_3S$, we find $T_c=264$~K.)

Finally, in the VII-substitution cases we have studied two cases: $\rm H_3S_{1-x}Cl_{x}$ and $\rm H_3Se_{1-x}Br_{x}$.
In general, the electronic energy bands shift down with respect to the Fermi energy, because of the increase of valence electrons;
the DOS at the Fermi energy, the phonon linewidths, and hence the electron-phonon coupling constants all decrease monotonously as the degree of VII-substitutions increases.
As a consequence, the $T_c$ drops monotonously with increasing the VII-substitution degree.
These results are summarized in Table~\ref{tab:dos} and Figs.~\ref{fig:tc}e-\ref{fig:tc}f (also see supplementary materials).

\begin{table}[b!]
\caption{\bf Fermi-surface DOS (in units of Hartree$^{\bf -1}$/spin) and electron-phonon coupling constant $\bf \lambda$ of $\rm \bf H_3S_{1-x}P_{x}$ under different pressures.}
\begin{tabular*}{7cm}{@{\extracolsep{\fill}}ccccccccc}
\hline\hline
\multirow{2}{*}{Pressure} & \multicolumn{2}{c}{H$_3$S$_{0.925}$P$_{0.075}$ } & \multicolumn{2}{c}{H$_3$S$_{0.9}$P$_{0.1}$}\\
\cline{2-5}
        & DOS  & $\lambda$ & DOS & $\lambda$\\
\hline   150~GPa& 9.16  &1.96& 9.13  &1.9\\
         175~GPa& 9.64  &2.0& 9.77  &1.98\\
         200~GPa& 9.92  &2.1& 10.55 &2.12\\
         225~GPa& 10.67  &2.24& 10.4 &2.19 \\
         250~GPa& 11.1  &2.44& 9.87 &2.24\\
\hline \hline
\end{tabular*}
\label{tab:pressdos}
\end{table}

\section{Discussion}

We have investigated the influence of partial atomic substitution on the superconductivity of hydrogen chalcogenides using first-principles calculations with VCA.
Our study has highlighted the key roles of strong covalent metallicity and low atomic mass in boosting the $T_c$ of BCS superconductivity.
The former can produce large electron-phonon couplings, whereas the latter can yield high-frequency phonon modes,
and in fact the highly-compressed $\rm H_3S$ constitutes both advantages.
We now take $\rm H_3S$ as the example to summarize our results.
In the VI-substitution cases, the reduction of DOS at Fermi energy dilutes the covalent metallicity
even though the coupling constant can be enhanced in the Te-substitution case, the $T_c$ is lowered because of the stronger atomic mass.
This leads us to further study the cases of Cl- and P-substitutions, in which the atomic mass remains hardly changed.
In the P-substitution case, the DOS, the phonon linewidths, the coupling constant, and hence the $T_c$ all increase as the substitution rate increases from zero up to $x=0.1$.
In sharp contrast, the oppose trend occurs in the Cl-substitution case.

In particular, we have shown that in the optimized case of $\rm H_3S_{0.925}P_{0.075}$ the $T_c$ may reach a record high value of $280$~K at $250$~GPa,
which is a feasible pressure in current experiments and would not induce any structure instability.
In light of our results, it might be possible that the silicon-substitution would also enhance the $T_c$ of high-pressure $\rm H_3S$ superconductor.
Because of the even less valance electrons, the optimum $T_c$ of $\rm H_3S_{1-x}Si_{x}$ might occur at even lower substitution concentration, which might be easier to realize in experiment.
For example, we find $T_c = 274$~K in the case of $\rm H_3S_{0.96}Si_{0.04}$ (see supplementary materials).
In the future, inclusion of anharmonic effects~\cite{Errea2015} may improve our predictions and thus better guide the experiments.
Nevertheless, our finding is exciting. It not only suggests that partial atomic substitution may lead to possible superconductivity above the ice point in the highly compressed $\rm H_3S$,
but also gives a hope that in principle low atomic mass and strong covalent metallicity may be designed in novel materials to realize high-$T_c$ BCS superconductivity under ambient pressures.

\section{Methods}

There are mainly two different procedures for disordered systems and partial atomic substitutions in first-principles calculations, i.e., an ordered supercell and the virtual crystal approximation (VCA)~\cite{Nordheim1931}. The former method is time-consuming and technically difficult to deal with the case for small concentrations. Thus, we chose to use VCA in this work. In calculations based on VCA, the primitive periodicity is retained but composed of virtual atomic potentials interpolating between the behaviors of actual components. Even though this approach neglects the local deformations around atoms and cannot explore the disordered structures very accurately, it often produces acceptable and useful results that have been verified in many research fields of condensed matter physics
\cite{Rosner2002,Boeri2004,Lee2004,Blackburn2011,Kimber2009,Bellaiche2002,Sani2004,Shevlin2005,Priour2005,Barbiellini2008,Noffsinger2009}.
The atomic substitution in the present work is simulated by the self-consistent VCA. For example, in H$_3$S$_{1-x}$Se$_x$ the virtual pseudopotentials of the S$_{1-x}$Se$_x$ is represented by a pseudopotential operator $V_{\rm VCA}= xV_{\rm Se}+(1-x)V_{\rm S}$, where $V_{\rm Se}$~($V_{\rm S}$) is the pseudopotential of Se (S) atom.

The present studies, including the electronic structures, the phonon spectra, and the electron-phonon couplings, were carried out using the ABINIT package~\cite{Gonze19971,Gonze19972,Gonze2005,Gonze2009} with the local-density approximation (LDA).
Hartwigsen-Goedecker-Hutter (HGH) pseudopotentials~\cite{Hartwigsen1998} were used in order to include spin-orbit couplings (SOC) for the heavy element, tellurium. The SOC for other elements were neglected since they are sufficiently light.
By requiring convergence of results, the kinetic energy cutoff of $30$~Hartree and the Monkhorst-Pack $k$-mesh of 40$\times$40$\times$40 were used in all calculations about the electronic ground-state properties. The phonon spectra and the electron-phonon couplings were calculated on a 8$\times$8$\times$8 $q$-grid.
Since calculating electron-phonon couplings referred to integrals over the Brillouin zone, we also carefully checked convergence of the results on the aforementioned $k$-mesh and $q$-grid, by comparing them with results in denser samples (a 40$\times$40$\times$40 $k$-mesh and a 10$\times$10$\times$10 $q$-grid).
\vspace{2mm}\\
\noindent{\bf Acknowledgements}
Y.F.G. and Y.G.Y were supported by the NSFC (Grants No.11174337 and No. 11225418),
the MOST Project of China (Grants No. 2014CB920903 and No. 2011CBA00100),
and the Specialized Research Fund for the Doctoral Program of Higher Education of China (Grant No. 20121101110046).
F.Z. was supported by UT Dallas research enhancement funds.
F.Z. acknowledges the Aspen Center for Physics (the NSF Grant No.1066293 and the Trustee's Fund) for hospitality during the finalization of this work.
F.Z. thanks Bing Lv for helpful discussions.
The computations were performed on TianHe-1(A) at the National Supercomputer Center in Tianjin.
\vspace{2mm}\\
\noindent{\bf Authors Contribution}
F.Z. and Y.G.Y. conceived the project. Y.F.G. carried out the calculations and data analysis.  Y.F.G, F.Z., and Y.G.Y. wrote the manuscript.
\vspace{2mm}\\
\noindent{\bf Competing Interests}
The authors declare that they have no competing financial interests.
\vspace{2mm}\\
\noindent{\bf Correspondence}
Correspondence and requests for materials should be addressed to Fan Zhang (zhang@utdallas.edu) and Yugui Yao (ygyao@bit.edu.cn).

\end{document}